\documentclass[]{gHPR2e}
\usepackage{graphicx}
\usepackage{amssymb}
\usepackage{amsmath}
\usepackage{color}

\advance\textheight by 1\baselineskip

\begin{document}
\doi{10.1080/0895795YYxxxxxxxx}
\issn{1477-2299}
\issnp{0895-7959} \jvol{00} \jnum{00} \jyear{2010} \jmonth{March}

\newcommand{\bk}{{\bf k}}
\newcommand{\bq}{{\bf q}}
\newcommand{\bQ}{{\bf Q}}
\newcommand{\bG}{{\bf G}}
\newcommand{\bK}{{\bf K}}
\newcommand{\bp}{{\bf p}}
\newcommand{\bx}{{\bf x}}
\newcommand{\by}{{\bf y}}
\newcommand{\br}{{\bf r}}
\newcommand{\bR}{{\bf R}}
\newcommand{\bJ}{{\bf J}}
\newcommand{\bz}{{\bf 0}}
\newcommand{\ba}{{\bf a}}
\newcommand{\bh}{{\bf h}}
\newcommand{\bd}{{\bf d}}
\newcommand{\bv}{{\bf v}}
\newcommand{\bdelta}{{\boldsymbol\delta}}
\newcommand{\Li}{{\mathop{\rm{Li}}\nolimits}}
\newcommand{\cotg}{{\mathop{\rm{cotg}}\nolimits}}
\renewcommand{\Im}{{\mathop{\rm{Im}}\nolimits\,}}
\renewcommand{\Re}{{\mathop{\rm{Re}}\nolimits\,}}
\newcommand{\sgn}{{\mathop{\rm{sgn}}\nolimits\,}}
\newcommand{\Tr}{{\mathop{\rm{Tr}}\nolimits\,}}
\newcommand{\EF}{E_{\mathrm{F}}}
\newcommand{\kB}{k_{\mathrm{B}}}
\newcommand{\kF}{k_{\mathrm{F}}}
\newcommand{\nF}{n_{\mathrm{F}}}
\newcommand{\Green}{{G}}
\newcommand{\dSC}{{\mathrm{dSC}}}
\newcommand{\dPG}{{\mathrm{dPG}}}
\newcommand{\dDW}{{\mathrm{dDW}}}
\newcommand{\Ret}{{\mathrm{R}}}
\newcommand{\Tau}{T_\tau}
\newcommand{\modified}[1]{{\relax #1}}

\markboth{Pellegrino, Angilella, Pucci}{Strain effect on the Drude weight of graphene}


\title{Effect of uniaxial strain on the Drude weight of graphene}

\author{F. M. D. Pellegrino$^{\rm a,b,c,d}$,
G. G. N. Angilella$^{\rm a,b,c,d,e}$$^{\ast}$\thanks{$^\ast$Corresponding author.
Email: giuseppe.angilella@ct.infn.it\vspace{6pt}}
and R. Pucci$^{\rm a,c}$\\\vspace{6pt}  
$^{\rm a}${\em{Dipartimento di Fisica e Astronomia, Universit\`a di Catania, Via
S. Sofia, 64, I-95123 Catania, Italy}}; $^{\rm b}${\em{Scuola Superiore di
Catania, Via S. Nullo, 5/i, I-95123 Catania, Italy}}; $^{\rm c}${\em{CNISM, UdR
Catania, I-95123 Catania, Italy}; $^{\rm d}${\em{CNR-IMM, Z.I. VIII Strada 5,
I-95121 Catania, Italy}}; $^{\rm e}${\em{INFN, Sez. Catania, I-95123 Catania,
Italy}};}\\\vspace{6pt}\received{\today}}

\maketitle

\begin{abstract}

We study the dependence on the strength and orientation of applied uniaxial
strain of the Drude weight in the conductivity of graphene. We find a
nonmonotonic dependence on strain, which may be related to the proximity to
several strain-induced electronic topological transitions. Given the relation
between the Drude weight and the long-wavelength plasmon frequency, such a
strain dependence can be evidenced by infrared spectroscopy measurements.\bigskip
\end{abstract}


\emph{Introduction.} The recent fabrication in the laboratory
\cite{Novoselov:05a} of graphene ---an atomically thick single layer of carbon
atoms in the $sp^2$ hybridization status--- has spurred a wave of research
activity both on the experimental and on the theoretical side. Graphene is
characterized by a honeycomb lattice, made of two interpenetrating triangular
sublattices. This gives rise to unusual electronic properties. The band
structure of graphene thus consists of two bands, touching at the Fermi level in
a linear, cone-like fashion at the so-called Dirac points $\pm\bK$. Indeed, the
low-energy quasiparticles in graphene follow much the same physics as
relativistic massless particles, thus allowing the observation of several
quantum-relativistic electrodynamic effects in a solid state system
\cite{CastroNeto:08}.

Equally remarkable are graphene's outstanding elastic properties. Despite its
quasi-two-dimensional character ---that should in principle even prevent its
very existence, owing to Mermin-Wagner's theorem \cite{CastroNeto:08}---
graphene is characterized by exceptional tensile strength and stiffness
\cite{Booth:08}, with a record breaking strength of $\sim 40$~N/m \cite{Lee:08},
which nearly equals the theoretical limit. The Young modulus of $\sim 1.0$~TPa
\cite{Lee:08} places graphene amongst the hardest materials known. Recent
\emph{ab initio} calculations \cite{Liu:07,Cadelano:09,Choi:10,Jiang:10} as well
as experiments \cite{Kim:09} have demonstrated that graphene single layers can
reversibly sustain elastic deformations as large as 20\%. It has been even
proposed that graphene-based nanodevices could be engineered, by tailoring the
desired electronic or transport properties on the strain-induced modifications
of the deformed graphene sheet (origami electronics) \cite{Pereira:09}. It has
been predicted that uniaxial strain should influence the optical conductivity of
graphene \cite{Pellegrino:09b} as well as its reflectivity in the optical range
\cite{Pellegrino:09c}.

Here, we will be concerned on the strain effects on the Drude weight of the
conductivity of graphene. This quantity can be connected by means of an
effective $f$-sum rule to the dispersion relation of plasmons \cite{Polini:09},
which has been studied also under applied strain \cite{Pellegrino:10a}.

\emph{Model.} The band structure of electrons residing on a honeycomb lattice
can be described in terms of the tight-binding Hamiltonian \cite{Pellegrino:09b}
\begin{equation}
H = \sum_{\bR,\ell} t_\ell a^\dag (\bR) b(\bR+\bdelta_\ell) + \mathrm{H.c.},
\label{eq:H}
\end{equation}
where $a^\dag (\bR)$ is a creation operator on the position $\bR$ of the A
sublattice, $b(\bR+\bdelta_\ell)$ is a destruction operator on a nearest
neighbor (NN) site $\bR+\bdelta_\ell$, belonging to the B sublattice, and 
$\bdelta_\ell$ are the vectors connecting a given site to its nearest neighbors,
their relaxed (unstrained) components being $\bdelta_1^{(0)} =
a(1,\sqrt{3})/2$,  $\bdelta_2^{(0)} = a(1,-\sqrt{3})/2$,  $\bdelta_3^{(0)} =
a(-1,0)$, with $a=1.42$~\AA, the equilibrium C--C distance in a graphene sheet
\cite{CastroNeto:08}. In Eq.~(\ref{eq:H}), $t_\ell \equiv t(\bdelta_\ell )$,
$\ell=1,2,3$, is the hopping parameter between two NN sites. In the absence of
strain they reduce to a single constant, $t_\ell \equiv t_0$, with $t_0 =
-2.8$~eV \cite{Reich:02}.

In terms of the strain tensor ${\boldsymbol\varepsilon} \equiv
{\boldsymbol\varepsilon}(\varepsilon,\theta)$ \cite{Pereira:08a} the deformed
lattice distances are related to the unstrained ones by $\bdelta_\ell =
(\mathbb{I} +  {\boldsymbol\varepsilon}) \cdot \bdelta^{(0)}_\ell$. Here,
$\theta$ denotes the angle along which the strain is applied, with respect to
the $x$ lattice coordinate, $\varepsilon$ is the strain modulus. The special
values $\theta=0$ and $\theta=\pi/6$ refer to strain along the armchair and
zig~zag directions, respectively.

The resulting tight-binding band structure features two bands $E_{\bk\lambda}$
($\lambda=1,2$), which close to the Fermi level can be approximated by Dirac
cones, touching at $\pm\bk_D$ \cite{Pellegrino:09b}. While at $\varepsilon=0$
these coincide with the high symmetry points $\pm\bK$ at the vertices of the
hexagonal first Brillouin zone (1BZ) of graphene, their positions move away from
$\pm\bK$ with increasing strain. Depending on $\varepsilon$ and $\theta$, they
may eventually merge into the midpoints of one of the sides of the 1BZ. In that
case, the cone approximation breaks down, and a finite gap opens at the Fermi
level \cite{Pellegrino:09b}. For intermediate strains, the constant energy
contours of $E_{\bk\lambda}$ can be grouped according to their topology, and are
divided by three separatrix lines \cite{Pellegrino:09b}. This corresponds to
having three distinct electronic topological transitions (ETT)
\cite{Lifshitz:60,Blanter:94,Varlamov:99}, which are here tuned by uniaxial
strain, as has been suggested to be the case in other quasi-2D materials, such
as the cuprates \cite{Angilella:01,Angilella:02d} and the Bechgaard salts
\cite{Angilella:02}. Correspondingly, while the density of states (DOS) is
characterized by a linear behaviour close to the Fermi level, because of the
linearity in $E_{\bk\lambda}$ at $\bk=\pm\bk_D$, a detailed analysis beyond the
cone approximation reveals the presence of logarithmic singularities in the DOS
at higher energies, that may be connected with the proximity to the ETTs
\cite{Pellegrino:09b}. This behaviour is reproduced in the frequency dependence
of the longitudinal optical conductivity $\sigma(\omega)$ and of the optical
reflectivity $R(\omega)$, which have been studied at increasing strain modulus
and different strain directions \cite{Pellegrino:09b,Pellegrino:09c}.

\emph{Results and conclusions.} We are interested in the longitudinal response
to an electric field forming an angle $\varphi$ with the $x$ crystallographic
direction. This is given by the dynamical conductivity \cite{Pellegrino:09b}
\begin{equation}
\sigma_{\varphi\varphi} (\omega) = \frac{ie^2 n}{m\omega} + \frac{ie^2}{\omega}
\lim_{q\to0} \Pi^{jj}_{\varphi\varphi} (\bq,-\bq,\omega),
\end{equation}
where $\Pi^{jj}_{\varphi\varphi} (\bq,-\bq,\omega)$ is the retarded
current-current ($jj$) correlation function, $e$ and $m$ are the electron charge and
mass, respectively, and $n$ is the carrier density. This is related to the
density-density ($\rho\rho$) correlation function $\Pi^{\rho\rho}$ through the
continuity equation. One obtains
\begin{equation}
\sigma_{\varphi\varphi} (\omega) = \frac{ie^2}{\omega} \lim_{q\to0}
\frac{\omega^2}{q^2} \Pi^{\rho\rho} (\bq,-\bq,\omega).
\end{equation}
Letting $\omega\to\omega+i0^+$, and extracting the real part, one recognizes the
Drude weight as
\begin{equation}
D_{\varphi\varphi} = \pi e^2 \lim_{q\to0} \frac{\omega^2}{q^2} \Re
\Pi^{\rho\rho} (\bq,-\bq,0).
\end{equation}
Within linear response theory, collective excitations are described by the
analytical properties of the appropriate correlation function. At lowest order,
this can be expressed as a polarization bubble. The random phase approximation
(RPA) is then the simplest diagrammatic procedure to include infinite orders in
the dielectric screening \cite{Giuliani:05}.
At low frequency and small wavevectors, the $\bk$-space integration involved in
the definition of $\Pi^{\rho\rho}$ can be performed analytically
\cite{Pellegrino:10a}. In that limit, the Fermi line in general reduces to an
ellipse of constant energy, $\mu$. One finds
\begin{equation}
D_{\varphi\varphi} = 4\mu\sigma_0 \left[\pi A_c^{-1} \rho_1 \left( \frac{\cos^2
(\varphi-\eta)}{A^2} + \frac{\sin^2 (\varphi-\eta)}{B^2} \right)\right] ,
\label{eq:Dpp}
\end{equation}
where $\sigma_0 = \pi e^2/2h$ is the so-called universal interband electrical
conductivity of neutral graphene \cite{Nair:08}, $A_c = 3\sqrt{3} a^2/2$ is the
area of the unit cell, $\rho_1$ is the strain-dependent prefactor in the linear
dependence of the DOS on the chemical potential at low energy, $\rho(\mu)=g_s
\rho_1 |\mu|$ \cite{Pellegrino:09b}, with $g_s =2$ for the spin degeneracy, and
$A$ and $B$ denote the semiaxes of the constant energy ellipse, with
$A^{-2} = \frac{1}{2} (\gamma - \sqrt{\alpha^2 + \beta^2} )$,
$B^{-2} = \frac{1}{2} (\gamma + \sqrt{\alpha^2 + \beta^2} )$;
$\alpha = -\frac{3}{2} a^2 (t_1^2 + t_2^2 - 2 t_3^2 )$,
$\beta  = \frac{3\sqrt{3}}{2} a^2 (t_1^2 - t_2^2 )$,
$\gamma = \frac{3}{2} a^2 (t_1^2 + t_2^2 + t_3^2 )$;
and
$\cos(2\eta) = |\alpha|/\sqrt{\alpha^2 + \beta^2}$,
$\sin(2\eta) = |\alpha|\beta/\alpha\sqrt{\alpha^2 + \beta^2}$.

In Eq.~(\ref{eq:Dpp}), the quantity between square brackets goes to unity in the
limit $\varepsilon\to0$, where in particular $\rho_1 = 2/(\pi\sqrt{3} t^2)$.
From Eq.~(\ref{eq:Dpp}), it follows that $D_{\varphi\varphi}$ attains its
maximum values whenever $\varphi -\eta = \pi/2$ (modulo $\pi$), and its minimum
values whenever $\varphi -\eta = 0$ (modulo $\pi$). It turns out that
$\eta=\theta$ (the angle defining the direction of applied strain) in the
zig-zag and armchair cases, whereas $\eta\simeq\theta$ in the generic case. 
Whilst Eq.~(\ref{eq:Dpp}) has been derived within the (deformed) cone
approximation, which implies low strain, experimental results as well as
\emph{ab initio} calculations \cite{Liu:07,Cadelano:09,Choi:10,Jiang:10} have
shown that both lattice and electronic structures of graphene are stable with
respect to strains as high as 20~\%. We therefore expect Eq.~(\ref{eq:Dpp}) to
be robust with respect to applied strain, within such strain ranges. In
particular, the main strain dependence is summarized by $\rho_1$, which
modulates the overall angular dependence of $D_{\varphi\varphi}$. Such a
quantity has been shown to increase steadily, and eventually diverge close to
the cone breakdown \cite{Pellegrino:09b}. Also the ellipse semiaxes $A$ and $B$
do depend on strain, whose role is that of increasing the ellipse anisotropy,
with $A=B$ when $\varepsilon=0$.
Such a dependence of the Drude weight on applied uniaxial strain is amenable to
experimental verification. In fact, the prefactor $\tilde{\omega}_1$ in the
long-wavelength dispersion relation of low-energy plasmons in graphene
\cite{Pellegrino:10a}, $\omega_\bq = \tilde{\omega}_1 \sqrt{qa}$, is related to
the Drude weight through \cite{Polini:09}
\begin{equation}
\frac{D_{\varphi\varphi}}{(\hbar\tilde{\omega}_1)^2 a} = 
\frac{2\pi\epsilon_0\epsilon_r}{\hbar} ,
\end{equation}
where $\epsilon_0$ is the dielectric constant in vacuum, $\epsilon_r =
(\epsilon_1 + \epsilon_2)/2$, with $\epsilon_1$, $\epsilon_2$ the relative
dielectric constants of the two media sandwiching the graphene sheet.

In conclusion, we have found a nonmonotonic strain dependence of the Drude
weight, related to the longitudinal polarization of graphene. This can be
described as an effect of the proximity to several strain-induced ETTs. Given
the connection between the Drude weight and the long-wavelength plasmon
frequency \cite{Polini:09,Pellegrino:10a}, such a dependence could be evidenced
by infrared spectroscopy \cite{Li:08}.

\bibliographystyle{gHPR}
\bibliography{a,b,c,d,e,f,g,h,i,j,k,l,m,n,o,p,q,r,s,t,u,v,w,x,y,z,zzproceedings,Angilella}
\end{document}